\documentstyle[twocolumn,aps,epsfig,prl]{revtex}
\begin{document}
\draft
\twocolumn[\hsize\textwidth\columnwidth\hsize\csname@twocolumnfalse\endcsname

\title{Identification of Electron Donor States in N-doped Carbon Nanotubes}

\author{ R. Czerw$^1$, M. Terrones$^{2,3}$, J.-C. Charlier$^4$, X. Blase$^5$, B. Foley$^{1,6}$, R. Kamalakaran$^2$,
N. Grobert$^7$, H. Terrones$^3$, \\P. M. Ajayan$^8$, W. Blau$^6$,
D. Tekleab$^1$, M. R\a"uhle$^2$, and D. L. Carroll$^1$}

\address{
             $^1$Dept. of Physics and Astronomy, Clemson University, Clemson SC 29634, USA\\
             $^2$Max-Planck-Institut f\"ur Metallforschung, Seestrasse 92, D-70174 Stuttgart, Germany\\
             $^3$Instituto de F\a'isica, Laboratorio Juriquilla, UNAM, A.P. 1-1010, 76000, Quer\a'etaro,
             M\a'exico\\
             $^4$Unit\a'e de Physico-Chimie et de Physique de Mat\a'eriaux, Universit\a'e Catholique de
             Louvain, Place Croix du Sud 1, B-1348, Louvain-la-Neuve, Belgium\\
             $^5$D\a'epartement de Physique des Mat\a'eriaux, Universit\a'e Claude Bernard, 43 bd. du 11 Novembre
             1918, 69622 Villeurbanne\\ Cedex, France\\
             $^6$Dept. of Physics, Trinity College Dublin, Dublin 2, Ireland\\
             $^7$Fullerene Science Center, CPES, University of Sussex, Brighton BN1 9 QJ, UK\\
             $^8$Dept of Materials Science and Engineering, Rensselaer Polytechnic Institute, Troy NY 12180-3590,
             USA}

\date{\today}
\maketitle

\begin{abstract}
Nitrogen doped carbon nanotubes have been synthesized using
pyrolysis and characterized by Scanning Tunneling Spectroscopy and
transmission electron microscopy.  The doped nanotubes are all
metallic and exhibit strong electron donor states near the Fermi
level. Using tight-binding and {\it ab initio} calculations, we
observe that pyridine-like N structures are responsible for the
metallic behavior and the prominent features near the Fermi level.
These electron rich structures are the first example of n-type
nanotubes, which could pave the way to real molecular
hetero-junction devices.
\end{abstract}


\vskip2pc]

Carbon nanotubes hold enormous promise in a wide variety of
electronic applications including, nano-heterojunctions
\cite{bla97,hu99,zha99}, diodes \cite{ant99}, interconnects
\cite{wei00,ter00,fuh00}, nano-transistors \cite{tan98,tan00},
sensors \cite{kon00,col98}, etc. The addition of dopants (e.g. N
or B) within the lattice of carbon nanotubes has suggested that
such applications might be realized. In this context, it has been
shown that B-doping of multiwalled carbon nanotubes (MWNTs)
results in the addition of acceptor states near the valence band
edge \cite{car98,bla99}. Further examples, such as doping through
`topological' defects, result in the introduction of electronic
defect states (e.g. by incorporating 5/7 defects via Stone-Wales
type transformations) \cite{cre97,hter00}.  These `dopants' have
also been theoretically predicted to result in acceptor and donor
states near the Fermi level \cite{cho00}. Although the exact
mechanisms and lattice effects of such doping schemes in carbon
nanotubes differ slightly to those of bulk materials such as Si,
the similarities to semiconductor physics are striking.  However,
the majority of the predicted and all of the experimentally
studied dopant effects result in modifications to the valence
band, and consequently the p-type conduction in carbon nanotubes.
This may be generally due to the nature of the {\it sp}$^2$-bonded
graphene lattice. Electron rich substitutions in such lattices
could easily lead to `of out of plane' bonding configurations,
which may induce curvature and closure of the system during growth
before tubular formation occurs.  For a complete analogy to bulk
semiconductor technology in low dimensional materials, we must
seek to dope the materials such that n-type and p-type conduction
occurs. Therefore, it is necessary to introduce donor states to
the system as well as acceptor states.

In this study we demonstrate that N-doping of the carbon nanotube
lattice, produced by pyrolytic routes, results in the
substitutional introduction of N into the C network.  From
tunneling spectroscopy studies, it is clear that the electronic
structure of these doped nanotubes was strongly modified by
including electron donor states near the conduction band edge. The
identification of the lattice `defects' was carried out by
tunneling spectroscopy and confirmed theoretically using tight
binding and {\it ab initio} calculations by comparing the
calculated local density of states (LDOS) with that obtained
experimentally. The results indicate that pyridine-like N
(two-bonded N) units embedded within the nanotube carbon lattice,
which differs from the direct substitution of three-fold
coordinated carbon, is responsible for a strong DOS in the
conduction band close to the E$_f$.

Multiwalled CN$_x$ carbon nanotubes/nanofibres were synthesized by
pyrolyzing ferrocene/melamine mixtures at 1050 $^\circ$C in an Ar
atmosphere. This technique has been recently described
\cite{ter99,han00}. The produced material consists of carpet-like
structures containing highly oriented nanotubes/nanofibres of
uniform diameter and length. HRTEM studies (JEM4000 EX operating
at 400 keV) of the products reveal the presence of nanofibres
exhibiting `bamboo-like' units, in which long and straight
segments (100-500 nm) of crystalline MWNTs are connected by
relatively `disordered' material.

\begin{figure}[!t]
\centerline{\epsfig{figure=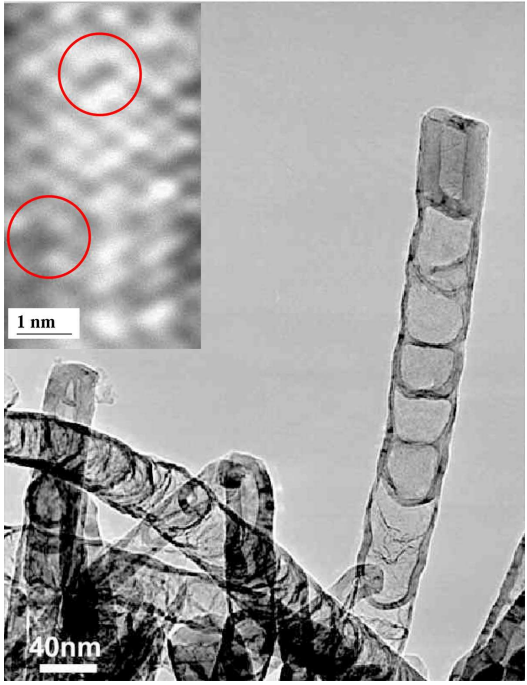,,width=0.8\linewidth,clip=}}\vspace{1
pc} \caption{TEM image of  typical N-doped fibers produced by
pyrolysis.  Note the `bamboo-like' structure consisting of
interconnected short (50-100 nm long) and hollow nanotube
segments.  (inset) Atomic resolution STM image of a small area on
the surface of a 20 nm diameter nanotube exhibiting distortions
and holes as discussed in the text.} \label{TEM}
\end{figure}

These `bamboo-type' structures are observed in Fig. 1 (TEM and
STM-inset). The walls of the `bamboo' segments exhibit some
disorder or distortion on the atomic scale (Fig. 1-inset). It is
noteworthy that the straight sections of the nanotube appear
crystalline in TEM with only small distortions along the outer
layers, while in STM large holes or gaps in the lattice are
clearly evident (circled region in Fig. 1-inset).  As will be seen
later, we attribute such holes to lattice discontinuities with
nitrogen decoration.  High resolution electron energy loss
spectroscopy (HREELS) studies of these and similar materials
reveal the incorporation of N within the graphene lattice at
levels between 2 \% and 10 \% (atom percent)
\cite{ter99,han00,mter99,mter98,sat99}.

For the STM study, flakes of the aligned carbon
nanotubes/nanofibres were dispersed ultrasonically in
tetrahydrofuran (THF). Exceptionally long ultrasonication was
necessary to breakup the tube bundles. However, TEM revealed
little (if any) damage resulting from this treatment.  This
solution was then drop cast onto freshly cleaved highly oriented
pyrolitic graphite (HOPG).  All STM was performed under ultra high
vacuum conditions ($<$ 10$^{-9}$ Torr) using mechanically formed
Pt - Ir tips. Stable imaging of the tubes was possible over an STM
setpoint range of +/- 500 mV @ 20 - 500 pA. Lattice images of HOPG
were used for X-Y calibration of the instrument while the step
height of Au was used for Z-calibrations. Images and spectra where
acquired at room temperature. STS was carried out concurrently
with image acquisition to insure tip placement on the tube
structure.  All STS was performed between +/-1.5 V to prevent
serious distortions or damage to the tube at higher voltages and
currents.  The spectra were acquired at a fixed gap distance for a
variety of setpoints.

Fig. 2 compares several tunneling spectra taken along a straight
section of the N-doped nanotube (Fig. 2a) with those of a standard
(pure) multiwalled carbon nanotube (MWNT) (Fig. 2b). All spectra
were converted to the equivalent local density of electronic
states (LDOS) using the accepted Feenstra algorithm of numerical
differentiation and normalization for tip height: (dI/dV)/(I/V).
It is important to note that all spectra analyzed in this work
were taken from areas over nanotubes that could be identified as
clean and free from `carbonaceous' and distorted surroundings. For
statistical completeness, several hundreds of spectra from ten
different tubes were recorded and compared.  The spectra shown
here are representative of this sampling and are for tubes that
exhibit similar outer shell diameter (ca. 20 nm). Notice that the
valence and conduction band features in the pure carbon tube (Fig.
2b) appear symmetric about the Fermi level, while for the N-doped
tube (Fig. 2a) an additional electronic feature occurs at $\sim$
0.18 eV. We also note that the presence of an electronic density
of states (DOS) at the E$_f$ indicates that the N-doped material
is metallic.  In addition, the electronic structure away from the
Fermi level is typical for MWNT's, exhibiting a variety of
symmetric features about the Fermi level (0 eV) that arise from
the one-dimensional nature of the nanotube and its inner shells.
The electron donor feature seen in the N-doped material is seen
everywhere along the straight sections of the nanotubes.  The
several spectra shown in Fig. 2 are taken on the same section but
at different locations, and they exhibit the N-doping feature
consistently at the same energy. This result is in contrast to the
B-doped case \cite{car98}, where variations in the peak position
are observed due to the formation of local phases, suggesting that
the N is distributed along the nanotube with small variations in
the N concentration (also observed experimentally in Ref. 18) but
no segregation into a separate phase.

There are several ways in which N can be incorporated in the
nanotube lattice \cite{miy97,her99}.  Earlier theoretical studies
\cite{yi93} have predicted donor states related to N-dopants
within carbon nantotubes at 0.27 eV below the bottom of the
conduction bands. However, these calculations were performed for
large band gap semiconducting single walled nanotubes (chiral and
small diameter tubes) and are not directly applicable here.
Another previous theoretical investigation \cite{cho00}
demonstrates that N placed substitutionally and fully coordinated
into the carbon lattice of a (10,10) single walled tube results in
quasibound electronic states approximately 0.53 eV from the Fermi
level.

\begin{figure}[!t]
\centerline{\epsfig{figure=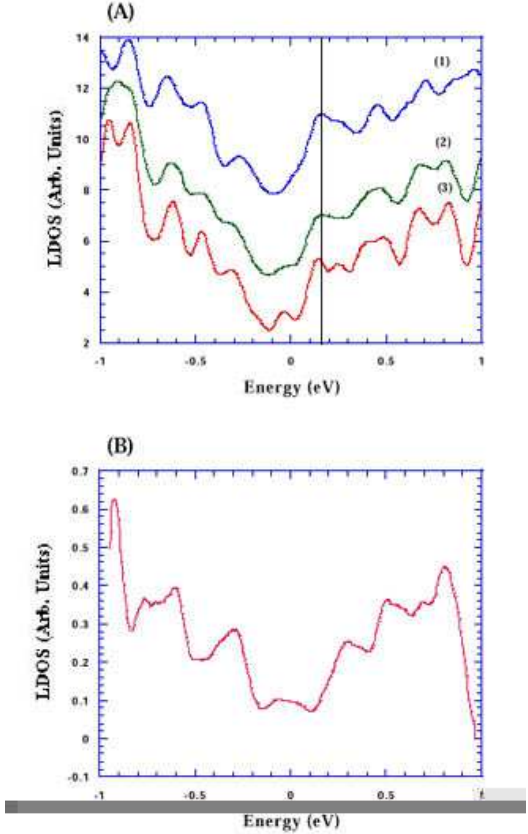,,width=0.8\linewidth,clip=}}\vspace{0.5
pc} \caption{(a) Tunneling spectra acquired on a straight section
of an N-doped carbon nanotube.  Spectra (1) - (3) were taken at
different locations along the surface but close to a hole as shown
in Fig. 1-inset.  Note the peak at 0.18 eV in all spectra.   (b)
Tunneling spectra along a pure MWNT of 20 nm. Notice the regularly
spaced van Hove singularities signifying the one-dimensional
nature of the material.} \label{LDOS}
\end{figure}

In the present study, we have performed {\it ab inito} band
structure calculations of N-doped graphitic systems, within the
local density approximation of density functional theory, in order
to simulate more accurately the doping effect on larger tube
surfaces \cite{non00} as measured here. A 3x3 graphite unit cell
was adopted, in which one of the 18 C atoms was replaced by a N
atom, leading to a $\sim$5.5\% substitutional doping. The main
effect of the N dopant in the `graphitic' DOS is the raising of
the Fermi level by $\sim$1.21 eV. There are indeed N-states
close-above the Fermi energy (in agreement with Ref. [16]).  On
general grounds, one may expect that the exact location of these
states will depend on doping level and local curvature of the
graphene sheet. However, these donor states, related to an
isolated N atom, are completely delocalized over several Angstroms
due to the metallic character of the undoped host structure, and
they are unable to explain the strong donor peak observed
experimentally in Fig. 2.

Careful experimental examination of the N-$\pi ^*$ edge in HREELS
reveals the formation of pyridine-type ({\it sp}$^2$-like)
structures within the lattice \cite{mter99}.  Therefore, an
alternative model for N-incorporation into the lattice might be
seen as that shown in Fig. 3 in which N rich cavities are formed
throughout the predominately `graphitic' network.  Further
evidence for such a model is observed in STM (Fig. 1-inset), which
reveals lattice `holes' and local atomic distortions along the
tube.  We observe that STM imaging of these materials proves to be
extraordinarily challenging as compared to pure MWNTs due to the
presence of local instabilities and lattice distortions.

\begin{figure}[!b]
\centerline{\epsfig{figure=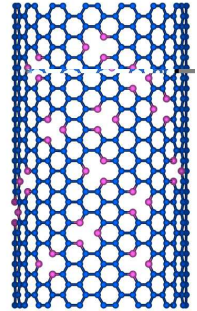,,width=0.3\linewidth,clip=}}\vspace{1
pc} \caption{Random N-doped (15,15) carbon nanotube with a
pyridine-like structure: The N is bonded to two carbon atoms (N:
red Spheres, C: blue spheres)} \label{Pyrid}
\end{figure}

In fact, the model consisting of pyridine-like units included in
carbon tubules results in a different electronic behavior from
those considered theoretically.  Specifically, the origin of the
low energy electronic states can be observed using tight binding
electronic structure calculations of the atomic arrangement as
shown in Fig. 3.  A simple recursion approach \cite{hay72}, which
has proved to capture the essential features associated with
composite B$_x$C$_y$N$_z$ nanotube heteojunctions \cite{bla97}, is
used to investigate the local DOS associated with the N dopants
arranged randomly and homogeneously in a pyridine-like
configuration within armchair and zigzag carbon nanotubes
\cite{tba}. Both DOS, represented in Fig. 4, exhibit prominent
donor peaks close-above the Fermi energy, in good agreement with
the experimentally determined peak positions of the N-doped tubes
(Fig. 2). As a consequence, N doping using pyridine-type units is
responsible for a strong related $\pi$ peak (shown in black, Fig.
4) just above the Fermi energy of the original undoped carbon
nanotube (red curves, Fig. 4). It is noteworthy that it does not
clearly result in an increase of the Fermi level of the host
carbon nanotube as in the case of an isolated N substituted in
graphite.

For SWNT's, the dopant states are quasibound states, which are 1-D
analogues of n-type donors in semiconductors. Since these states
are resonant with the continuum of conducting states in the 1-D
case, the local conductance of the tube near the defect area drops
\cite{cho00}.  In the N-doped MWNT case, the observed `bamboo'
structure will almost certainly be the dominate scattering
mechanism in the tube. Tunneling spectroscopy over regions that
could be clearly identified as interconnections between the
nanotube sections show that these regions are also highly
conductive (not shown).

\begin{figure}[!b]
\centerline{\epsfig{figure=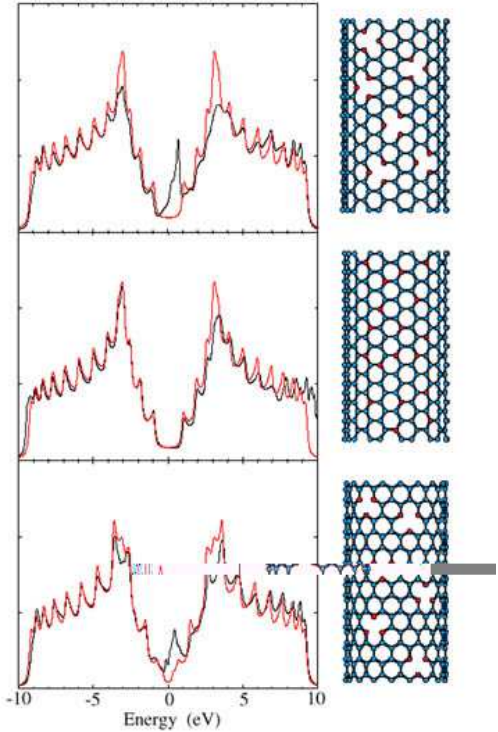,,width=0.8\linewidth,clip=}}\vspace{1
pc} \caption{Theoretical LDOS associated with a Pyridine-like
structure of N-doping in (top) armchair (10,10) and (bottom)
zigzag (17,0) nanotubes. In both cases, a random, but homogeneous
doping (N: red spheres - C: blue spheres) is adopted as
illustrated on the right-hand side of the figure. The LDOS of
doped (black curve) and pure (red curve) carbon nanotubes are
compared. Such Pyridine-like structure of N-doping is responsible
for the prominent donor-like features close-above the Fermi
energy. Note that the semiconducting (zigzag) nanotube becomes
metallic after introducing N in the carbon lattice.}
\label{Pyridine}
\end{figure}

The results presented here also provide the first evidence of the
synthesis of local domains of the predicted planar C$_3$N$_4$ or
CN \cite{miy97,liu89,tet96} within MWNT's (exhibiting the topology
of our pyridine-like structures). However, our CNx structures are
not crystalline in three dimensions and appear to be discontinuous
along the tube surface, which may also explain the fact that all
tubular structures are metallic.  However, the mechanical
properties of these tubes may be important in the fabrication of
composites of high tensile strength \cite{her99} and shock
absorbers.

In summary, we have demonstrated that N-doping of carbon nanotubes
leads to the introduction of conduction band modifications
including a large electron donor state.  This narrow state lies
approximately 0.2 eV from the Femi level.  The local environment
of the N within the carbon network mainly consists of N-C
structures arranged in a pyridine-like configuration, which also
explains the metallic behavior observed in these nanostructures.
Our calculations reveal that electronic signatures of
pyridine-like units are relatively insensitive to tube chirality
or the proximity between these rings. This novel doping scheme not
only supplies us with a road map for insertion of other electron
rich impurities in the carbon lattice for the creation of donor
states but may well suggest a way to achieve full p-n junctions in
carbon nanotubes. Interestingly, we note that connections between
B-doped carbon nanotubes reported elsewhere
\cite{car98,bla99,her00} and these materials should result in a
barrier of approximately 0.5 V.

We are grateful to AFOSR (DLC), the NSF (DLC), the Alexander von
Humboldt Stifftung (MT), CONACYT-M\a'exico grants: 25237 and
J31192U (HT, MT), DGAPA- UNAM grant 108199 (HT), the Royal Society
(NG) and the Max-Planck-Gesellschaft (RK). JCC acknowledges the
National Fund for Scientific Research [FNRS] of Belgium and the
Belgian Program on Inter-university Attraction Poles on Reduced
Dimensionality Systems (PAI4/10). Part of this work is carried out
within the framework of the EU COMELCAN contract
(HPRN-CT-2000-00128). PMA acknowledges the National Science
Foundation for supporting his research through the CAREER program.


\begin{references}
\bibitem{non00}
Nonlocal pseudopotentials are generated for C and N atoms with a
36 Ry energy cutoff for the plane wave expansion of the
wavefunctions.
\bibitem{tba}
The tight-binding Hamiltonian is restricted to 2p electrons ($\pi$
and $\pi ^*$ orbitals only), and the parameters are chosen so as
to reproduce the correct band gap for BN and BC$_2$N nanotubes and
the acceptor and donor levels for substitional C (B or N) atoms in
BN-based (C-based) systems [see Ref.12].
\end{references}
\end{document}